\documentclass[aps,preprint,showpacs,floatfix,superscriptaddress]{revtex4}

\usepackage{graphicx}

\begin{document}

\title{Chiral charge order in the superconductor 2H-TaS$_2$}

\author{I. Guillam\'on}
\affiliation{Laboratorio de Bajas Temperaturas, Departamento de F\'isica de la Materia Condensada, Instituto de Ciencia de Materiales Nicol\'as Cabrera, Facultad de Ciencias, Universidad Aut\'onoma de Madrid, E-28049 Madrid, Spain}
\affiliation{H.H. Wills Physics Laboratory, University of Bristol, Tyndall Avenue, Bristol BS8 1TL, UK}
\author{H. Suderow}
\affiliation{Laboratorio de Bajas Temperaturas, Departamento de F\'isica de la Materia Condensada, Instituto de Ciencia de Materiales Nicol\'as Cabrera, Facultad de Ciencias, Universidad Aut\'onoma de Madrid, E-28049 Madrid, Spain}
\author{J. G. Rodrigo}
\affiliation{Laboratorio de Bajas Temperaturas, Departamento de F\'isica de la Materia Condensada, Instituto de Ciencia de Materiales Nicol\'as Cabrera, Facultad de Ciencias, Universidad Aut\'onoma de Madrid, E-28049 Madrid, Spain}
\author{S. Vieira}
\affiliation{Laboratorio de Bajas Temperaturas, Departamento de F\'isica de la Materia Condensada, Instituto de Ciencia de Materiales Nicol\'as Cabrera, Facultad de Ciencias, Universidad Aut\'onoma de Madrid, E-28049 Madrid, Spain}
\author{P. Rodi\`ere}
\affiliation{Institut N\'{e}el, CNRS / UJF, 25, Av. des Martyrs, BP166, 38042 Grenoble Cedex 9, France}
\author{L. Cario}
\affiliation{Institut des Mat\'eriaux Jean Rouxel (IMN),  Universit\'e de Nantes - CNRS, 2 rue de la Houssini\'ere, BP 32229, 44322 Nantes Cedex 03, France}
\author{E. Navarro-Moratalla}
\affiliation{Instituto de Ciencia Molecular (ICMol), Universidad de Valencia, Catedr\'atico Jos\'e Beltr\'an 2, 46980 Paterna, Spain}
\author{C. Mart\'i-Gastaldo}
\affiliation{Instituto de Ciencia Molecular (ICMol), Universidad de Valencia, Catedr\'atico Jos\'e Beltr\'an 2, 46980 Paterna, Spain}
\author{E. Coronado}
\affiliation{Instituto de Ciencia Molecular (ICMol), Universidad de Valencia, Catedr\'atico Jos\'e Beltr\'an 2, 46980 Paterna, Spain}

\begin{abstract}
We find chiral charge order in the superconductor 2H-TaS$_{2}$ using Scanning Tunneling Microscopy and Spectroscopy (STM/S) at 0.1 K. Topographic images show hexagonal atomic lattice and charge density wave (CDW) with clockwise and counterclockwise charge modulations. Tunneling spectroscopy reveals the superconducting density of states, disappearing at T$_c$ = 1.75 K and showing a wide distribution of values of the superconducting gap, centered around $\Delta$=0.28 meV.
\end{abstract}


\maketitle

Since the suggestion by Anderson and Blount\cite{Anderson65} that charge order may appear in metals and superconductors, the size and anisotropy of charge density wave (CDW) modulations are an intensely debated issue\cite{Kiss07,Rossnagel01}. Layered compounds, such as hexagonal bilayer stackings of dichalcogenides, named 2H, have received attention because the charge order is only partial and clean metallic behavior is ubiquitously observed.  In the 2H series of TaSe$_2$, TaS$_2$, NbSe$_2$ and NbS$_2$, the critical temperature of the CDW decreases from around 120 K in 2H-TaSe$_2$, to 80 K in 2H-TaS$_2$, down to 30 K in 2H-NbSe$_2$ and finally no CDW in 2H-NbS$_2$ \cite{CastroNeto01,Suderow05d}. Superconductivity coexists with CDW, with a superconducting critical temperature T$_c$ which increases from around 0.2 K in 2H-TaSe$_2$ up to 7.2 K and 6 K in, respectively 2H-NbSe$_2$ and 2H-NbS$_2$. Generally it is believed that their mutual interaction is competitive, as charge order usually drives towards insulating behavior, and superconductivity towards perfect conductance. However, evidence for cooperative interactions, which actually shape the form of the gap over the Fermi surface, have been found in angular resolved photoemission \cite{Kiss07}. CDW significantly shapes the anisotropy of the superconducting properties of the compound 2H-NbSe$_2$, producing vortex cores with a star shaped cross section, instead of the tubular cores expected in a simple isotropic s-wave superconductor\cite{Hess90,Gygi90,Hayashi96,Guillamon08}.

The Fermi surface of all four 2H compounds (2H-TaSe$_2$, 2H-TaS$_2$, 2H-NbSe$_2$ and 2H-NbS$_2$) consists essentially of warped tubes deriving from the d electrons of the transition metal, with a strong hybridization to the p electrons of the chalcogen\cite{Wilson75,Corcoran94}. Calculations find that, generally, p-electron derived bands of the chalcogens do not cross the Fermi level, with the important exception of 2H-NbSe$_2$, where they form a small 3D pocket\cite{Johannes06,Ge10}. The charge order properties have been studied in the whole series, although usually at temperatures well above liquid helium (see e.g. \cite{Sacks98}). Only the superconducting properties of 2H-NbSe$_2$ and of 2H-NbS$_2$ have been analyzed in some detail, finding peculiar multigap superconducting behavior \cite{Guillamon08,Boaknin03,Rodrigo04c,Fletcher07,Noat10,Kacmarcik10}.

On the other hand, in the related semimetallic compound 1T-TiSe$_2$ (the 1T trigonal arrangement of single layers produces generally semimetallic or semiconducting systems \cite{Bulaevskii76,DiSalvo76}), CDW has been found with novel properties. 1T-TiSe$_2$ presents the hitherto unique feature of alternating sheets with a 3D helical stacking of intralayer charge modulation with the helical axis perpendicular to the layers\cite{Ishioka10}. Domains which are mirror images of each other, and cannot be brought together with only rotation operations, are found over the surface. The question if such kind of chiral charge order can appear in a superconductor with CDW has not been addressed. Chiral symmetry is a particular example of absence of space parity, a concept of fundamental importance
in BCS theory. To our knowledge, the only known superconductor where chiral symmetry is presently debated is the p-wave superconductor Sr$_2$RuO$_4$\cite{Mackenzie03}. In Sr$_2$RuO$_4$, the chiral symmetry appears due to Cooper pair wavefunction phase winding at the superconducting transition, related to the particular form of the p-wave order parameter. How superconductivity evolves out of a chiral electronic environment, however, has yet to be worked out. Here we find for the first time chiral charge order in a metallic compound, 2H-TaS$_2$, which, in addition, superconducts below 1.75K. We discuss chiral properties of CDW in this material, and determine the superconducting density of states from tunneling spectroscopy.

2H-TaS$_2$ single crystals were synthesised using a iodine transport method. Stoichiometric amounts of the elements were sealed under vacuum in a silica tube with a small amount of Iodine (5mg/cm$^3$). The tube was then heated  for a period of 15 days in a gradient furnace ($\approx$50$^{\circ}$C along the tube) with the mixture located in the high temperature zone ($\approx$ 800$^{\circ}$C). This synthesis yielded layered crystals some hundreds of $\mu$m size that exhibit X-ray diffraction patterns corresponding to 2H-TaS$_2$. For Scanning Tunneling Microscopy (STM) measurements, we use a STM in a partially home-made dilution refrigerator set-up described in detail elsewhere \cite{Rodrigo04b,Suderow11b}. The set-up features the possibility to mount several very small (some tens of micron size) single crystals and change the position of the tip from one sample to the other in-situ, at low temperatures. The gold tip can be moved to a sample of gold to clean and sharpen, and the energy resolution of the set-up is of 15$\mu$eV. Scanning was made at constant current with tunneling conductances of some 0.1 $\mu$S and bias voltages usually below 10 mV. Several 2H-TaS$_2$ samples were glued one by one using silver epoxy on an Au substrate, and measured at 100 mK. Samples were mechanically cleaved using scotch tape prior to the experiment. After cleaving, clean atomically flat surfaces are easily found, with the exposed surface composed of hexagons of the chalcogen atomic layer.

\begin{figure}[ht]
\includegraphics[clip=true,width=16cm]{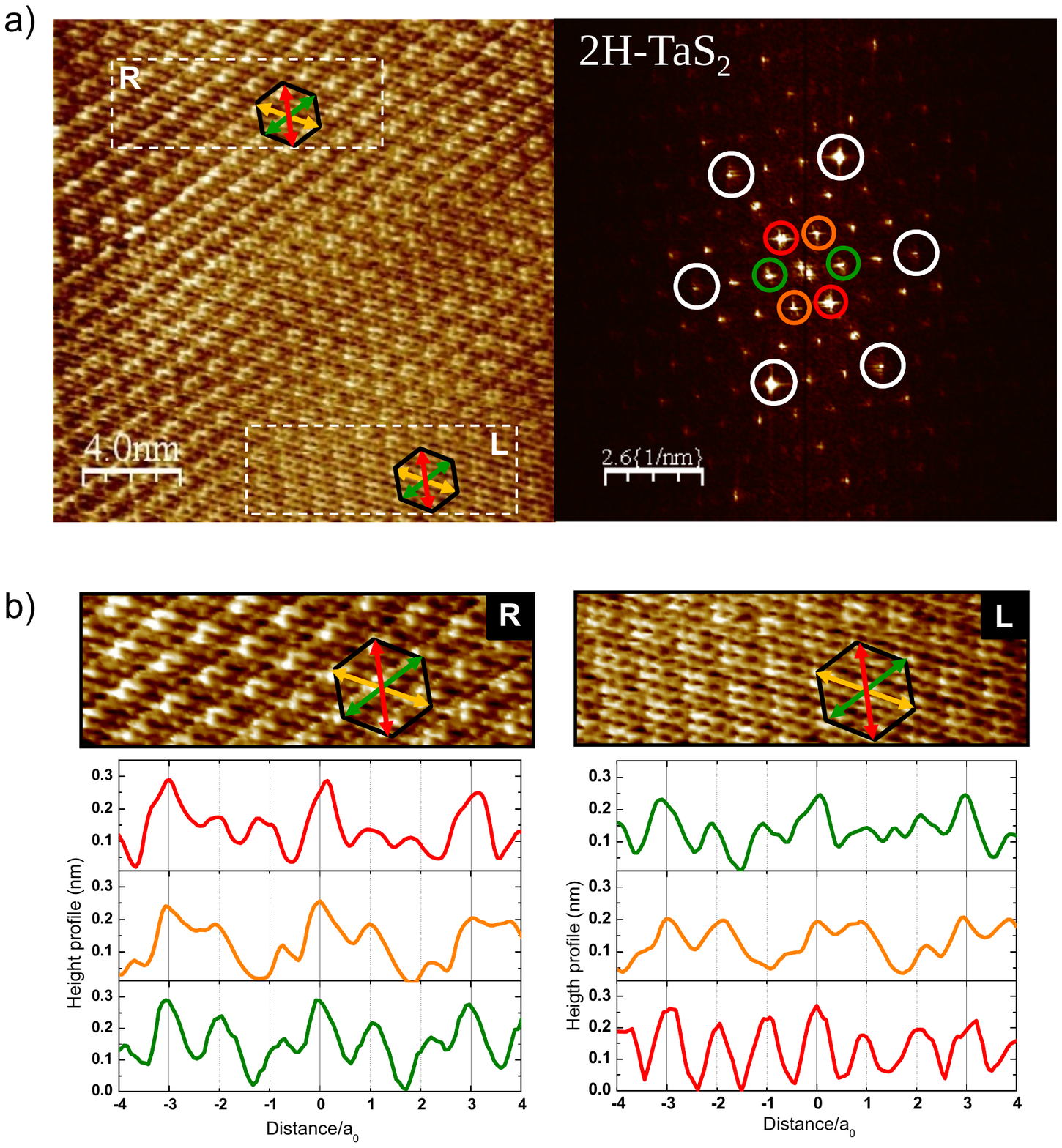}
\vskip -0 cm
\caption{{\small{In a we show the STM topography at constant current made at 5 mV in 2H-TaS$_2$ at 100 mK (left panel), and its corresponding Fourier transform (right panel). White circles show Bragg peaks corresponding to the atomic S lattice, and colored circles those corresponding to the three CDW with reciprocal lattice vectors $q_{CDW_1}$ (red), $q_{CDW_2}$ (orange) and $q_{CDW_3}$ (green). In b we show real space profiles taken along the three principal directions of the CDW modulations as marked by the arrows in a) and zoomed-up images in upper panels (corresponding to areas marked by dashed white rectangles in the top (R) and bottom parts (L) of the image). It gives a peculiar Moir\'e pattern where the height relationship changes from a rightwise rotating angular dependence to a leftwise rotating angular dependence in the same figure, without any defect or step observed in between.}}}
\end{figure}

The STM topography at constant current of 2H-TaS$_2$ taken at 100 mK is shown in Fig.1a. The hexagonal atomic S lattice is nicely observed, together with a remarkable triangular 3x3 charge modulation of wavelength 3 times the lattice constant. The Fourier transform of the whole image (right panel of Fig.1a) shows that the CDW modulation is located at one third the in-plane reciprocal lattice wavevectors, within experimental error of $\pm$ 1\%. The CDW modulation leads to hexagons spanned by 3 vectors parallel to the three symmetry axis of the hexagonal atomic lattice, repeating itself in the topography. We identify two regions with different hexagonal patterns produced by the CDW modulation, marked by white rectangles in Fig.1. In each region, within a given hexagon, the amplitude of the charge density along the three principal CDW vectors has an angular dependence, breaking six-fold rotational symmetry. The amplitude relationship between the three modulations varies when going from the upper left corner to the lower right corner. In Fig.1b we show representative topography profiles along the three CDW wavevectors. The CDW amplitude is highest along the red arrow direction in the upper left corner, and along the green arrow direction in the lower right corner. The hexagonal pattern spanning the CDW modulation, changes as a function of the position in the image, without any step or impurity cutting the image in two. Note that both parts cannot be transferred into each other by a simple rotation.

\begin{figure}[h]
\begin{center}
\includegraphics[clip=true,width=11cm]{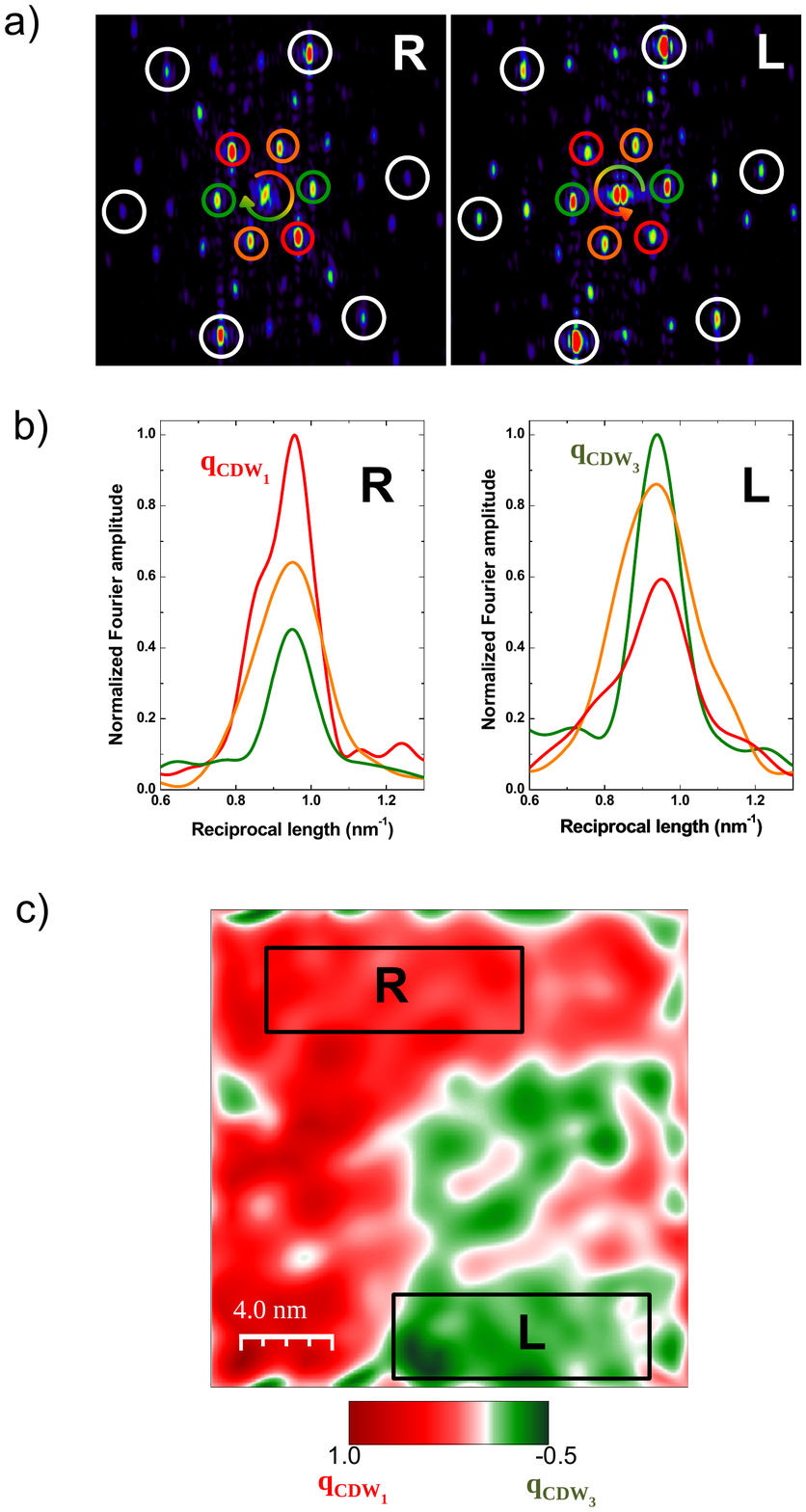}
\end{center}
\vskip -1.2 cm
\caption{{\small{Fourier analysis of the CDW modulations of Fig.1. In a we show the Fourier transform of the rectangles in the top (R) and bottom (L) areas of Fig.1a. Red, orange and green circles mark the Bragg peaks corresponding to each CDW modulation direction. The intensity of the peaks decreases, respectively, clockwise (R, from red to green in b) and anticlockwise (L, from green to red in b). In b, profiles are normalized to one at the highest peak. In c we show Fourier amplitude map highlighting the spatial distribution of the dominant CDW components. The map is obtained as described in the text and shows the difference between the amplitudes of $q_{CDW_1}$ and $q_{CDW_3}$ normalized by the highest value of the $CDW_1$ amplitude. The area where $q_{CDW_1}$ has highest amplitude is marked in red, and the area where $q_{CDW_3}$ is highest is marked in green. White shows the boundary, where the size of both modulations is equal.}}}
\end{figure}

The angular dependence of the CDW modulation observed in the real space can be further characterized by making a detailed Fourier analysis. If we look at the Fourier transforms of the two different areas marked by rectangles in Fig.1a, we find that the amplitude of CDW Bragg peaks decrease clockwise (in R) and counter-clockwise (in L), respectively. The normalized amplitude relationships for, respectively, $q_{CDW_1}$ (red), $q_{CDW_2}$ (orange), and $q_{CDW_3}$ (green), are 1-0.6-0.4 for the upper part of the image (R in Fig.2b), and 0.6-0.8-1 for the bottom part of the image (L in Fig.2b). This behavior is not followed by the intensity changes of the lattice Bragg peaks and evidences the presence of chiral domains in the CDW of 2H-TaS$_{2}$.

The spatial distribution of the chiral domains can be imaged by further Fourier analysis. We first Fourier transform the whole real space image (shown in Fig.1a), center the resulting Fourier image at the peak corresponding to CDW's component with highest amplitude in the upper left corner domain (q$_{CDW_1}$), filter the rest of the image out, and re-transform into real space. This gives us the real space dependence of the CDW amplitude corresponding to the wavevector q$_{CDW_1}$. We then make the same procedure for the CDW modulation corresponding to q$_{CDW_3}$  which has the highest amplitude in the lower right corner domain. Finally, we subtract both images\cite{Zeljkovic11} obtaining the map of the relative height between both modulations (Fig.2c). We use red color for the area where q$_{CDW_1}$ has highest amplitude, and green color for the area where q$_{CDW_3}$ has highest amplitude (Fig.2c). White color is used for the boundary, where both modulations have equal amplitude. The size of the areas with different CDW modulations is rather large, of up to 10 nm, and their boundary is irregular.

The height of the atomic and CDW Fourier Bragg peaks measured in the transition metal di-chalcogenides depends always on the direction. This is the result of the complex Moir\'e-pattern, with very different hexagonal symmetry structures appearing in different positions, ubiquitously observed in these materials (see e.g. \cite{Coleman88,Giambattista88}). Calculating precisely the local density of states to obtain such patterns is a problem of great complexity. Tip-sample coupling, depending on the detailed band structure of the sample, and of the form and geometry of the tip, or defects, shape considerably the measured local density of states \cite{Tersoff83,Guillamon08PRB}. For example, Moir\'e patterns of 2H-NbSe$_2$ have been widely discussed in literature, and related to different tip-sample coupling effects. In 2H-TaS$_2$, within the same image, without any change in the properties of the tip nor of the surface (i.e. without any step, defect, or surface anomaly), the Moir\'e patterns change its symmetry properties, in a rather complex way, as shown in Figs. 1 and 2. This was first found in 1T-TiSe$_2$ \cite{Ishioka10}, and, to the best of our knowledge, was never observed previously in any metallic 2H dichalcogenide compound. To explain such patterns, it is needed to consider out-of-plane decoupling of the three components of the CDW, which are combined in two different ways in different areas of the surface, giving chiral CDW order\cite{Ishioka10}.

\begin{figure}[ht]
\begin{center}
\includegraphics[clip=true,width=13cm]{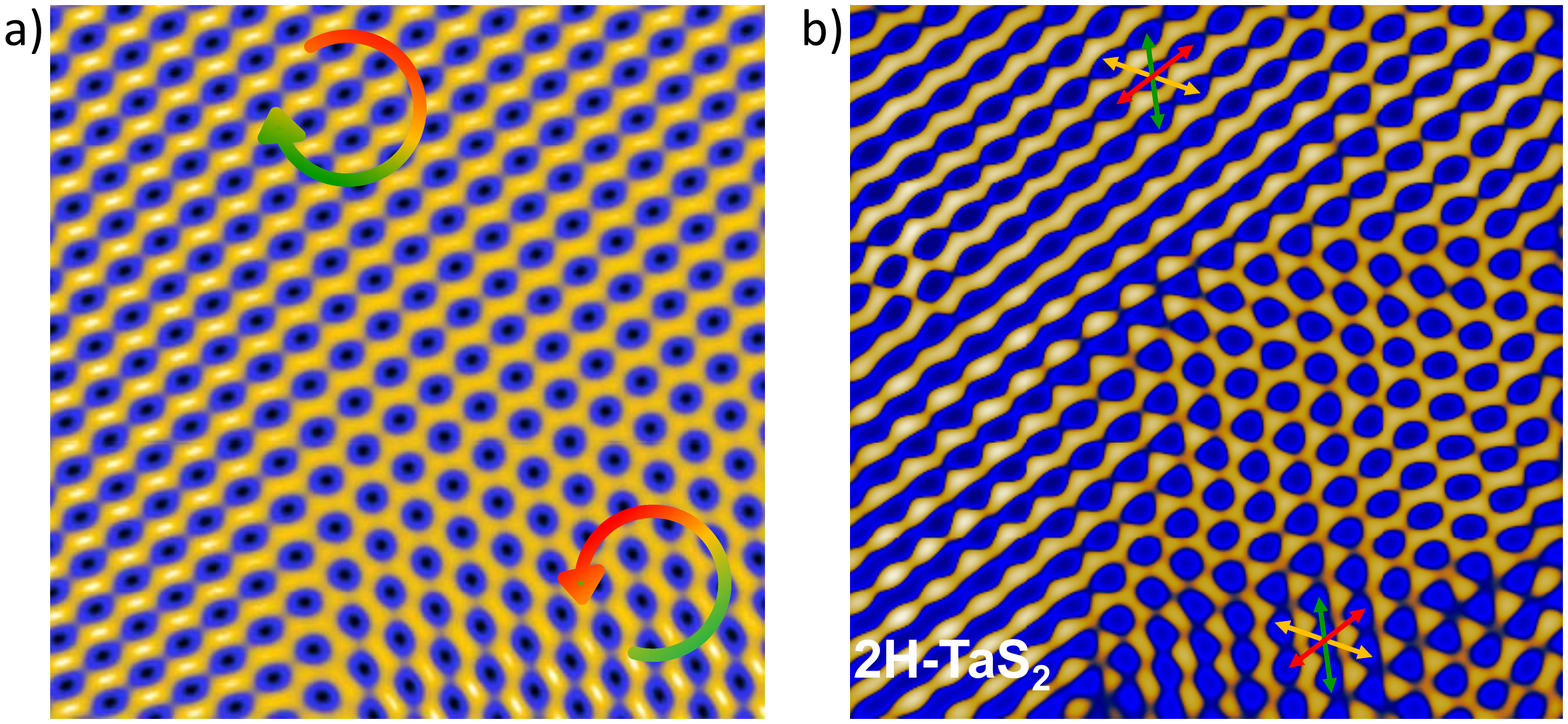}
\end{center}
\vskip -1,0 cm
\caption{{\small{In a) we show the charge modulation obtained by summing up three $CDW_i=A_icos($\textbf{$q_{CDW_i}$} \textbf{r}$)$, using $A_1=1,A_2=0.6,A_3=0.3$ in the upper left corner's chiral domain (clockwise decrease of the amplitude of the CDW Fourier peaks as shown by the arrow), and $A_1=0.3,A_2=0.6,A_3=1$ in the lower right corner's chiral domain (counterclockwise decrease of the amplitude of the CDW Fourier peaks as shown by the arrow). These modulations give differently oriented oval shaped patterns. In between, the difference between the $A_i$ are smoothly varied as a function of the position. As the difference in $A_i$ becomes smaller, a circularly shaped pattern is visible. b) shows the CDW modulation observed in the experiment. Atomic lattice modulations are removed from the Fig.1 by Fourier transform filtering. Oval shaped patterns are observed in the two chiral domains, with more round circular patterns along the chiral domain wall. Arrows indicate main CDW modulation directions, as in Fig.1. In both figures, the color goes continously from dark blue for highest charge density to white for lowest charge density.}}}
\end{figure}

Microscopic origin of chiral charge arrangements is now under debate. Electronic interactions between planes (along c), which separate in-plane charge modulations through Coulomb interaction, can lead to a c-axis chiral modulation of the CDW\cite{Ishioka10}. More recently, it has been shown that chiral charge order can also appear in relation to orbital order, which produces transversal polarization introducing phase shifts in the CDW \cite{Jasper11}. The charge modulation observed in 2H-TaS$_2$ can be simulated by assuming that the intensity of the modulations corresponding to each $CDW_i$ changes as a function of the out-of-plane stacking, or by phase shifting each component \cite{Ishioka10,Jasper11}. For example, if we sum up three CDW components with $CDW_i($\textbf{r})$=A_i cos($\textbf{$q_{CDW_i}$} \textbf{r}$)$ (with A$_i$ the amplitude of \textbf{$q_{CDW_i}$}), and two different amplitude relationships in two different areas, we obtain the image shown in Fig.3a. Qualitatively, the obtained shapes agree with the experimental image of the CDW (Fig.3b).

All these observations, which are similar to those reported in the compound 1T-TiSe$_2$\cite{Ishioka10}, demonstrate that 2H-TaS$_2$ has a chiral modulation of its CDW. Note that 2H-TaS$_2$ is a metal which superconducts, but 1T-TiSe$_2$ is a semimetal, with no report of superconducting correlations at ambient pressure\cite{Kusmartseva09}. Note also that in 1T-TiSe$_2$ the CDW modulation is located at two times the lattice constant, whereas in 2H-TaS$_2$ it is located at three times the lattice constant. The three fold in-plane symmetry with clockwise and counterclockwise domains appearing in the charge order features, are, however, similar. Furthermore, in 1T-TiSe$_2$, there is a c-axis modulation of the CDW \cite{Ishioka10}, with periodicity 2 times the c-axis lattice constant. In 2H-TaSe$_2$, interlayer separation is highest among the 2H series (it is continuously decreasing from 2H-TaSe$_2$, 2H-TaS$_2$, 2H-NbSe$_2$ and is smallest in 2H-NbS$_2$ where no CDW is observed) the three coexisting CDWs lead to a reduction of the crystal symmetry from hexagonal to orthorhombic \cite{Littlewood82}. In 2H-NbSe$_2$, as discussed above, no chiral CDW has been reported, in spite of having been very extensively studied by many different groups. It could be that 2H-TaS$_2$ lies in between both cases, being sufficiently 2D to present de-coupling of CDW modulations, but not enough to show structural transition or to destroy superconductivity through a too strong destruction of the Fermi surface by the CDW gap opening.

The superconducting tunneling conductance of 2H-TaS$_2$ measured at 100 mK is shown in Fig.4. The tunneling conductance remains flat until around 0.15 mV, where it starts increasing continuously until 0.4 mV, where the quasiparticle peak is located. Such a feature is found over large parts of the surface, and we could not, until now, observe changes related to the domains of the chiral CDW modulation. The behavior is instead reminiscent of the compounds 2H-NbSe$_2$ and 2H-NbS$_2$, where a distribution of values of the superconducting gap, starting also roughly at half the maximal gap value, is found. In the latter compounds, however, two clear features are seen in the tunneling conductance, which show up as two peaks in the derivative of the corresponding density of states\cite{Guillamon08,Rodrigo04c}. These features are due to the two band superconducting properties characteristic of these compounds. Here, the derivative of the density of states (DOS, obtained by de-convoluting the temperature, as shown in Ref.\cite{Guillamon08}), shows only one broad peak centered at $\Delta$=0.28 meV (Fig.4), which evolves until the T$_c$=1.75 K following roughly BCS theory. The value we find for $\Delta$ is close to, although somewhat larger, than the value expected within BCS theory, $\Delta$=1.76k$_B$T$_c$=0.265 meV. Thus, the superconducting gap of 2H-TaS$_2$ has a significant anisotropy, and the two gap features characteristic of 2H-NbSe$_2$ and 2H-NbS$_2$ are, if they exist at all, less clearly resolved in the superconducting density of states.

\begin{figure}[ht]
\begin{center}
\includegraphics[clip=true,width=14cm]{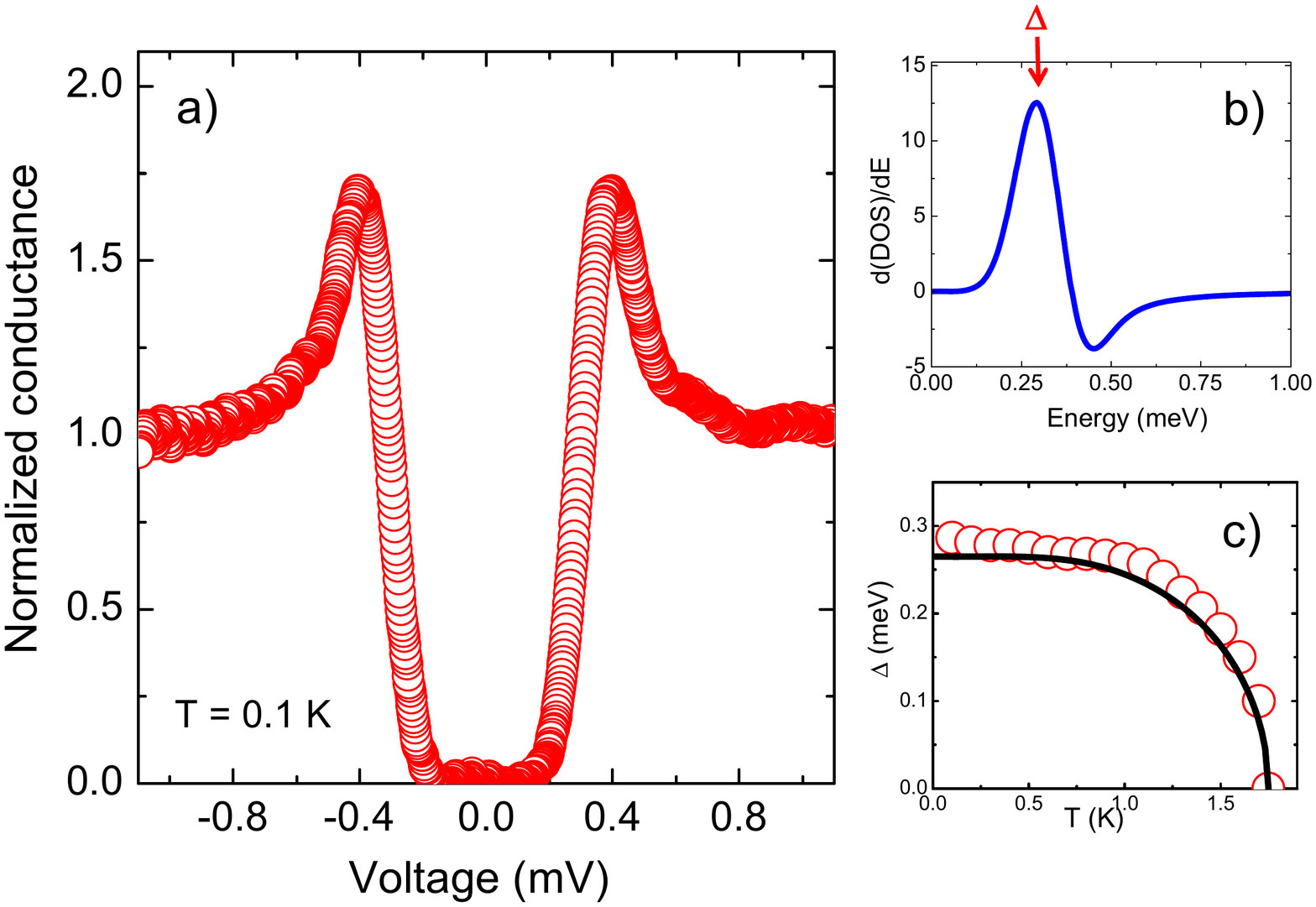}
\end{center}
%\vskip -4 cm
\caption{{\small{a. Normalized tunneling conductance curve obtained in 2H-TaS$_2$ at 0.1 K (conductance $\sigma$ at 1mV is of 1.7 $\mu$S). In b we show the derivative of the associated local density of states (DOS), which has been obtained by de-convoluting the temperature from the tunneling conductance. The derivative has a peak at $\Delta=$0.28 meV. In c we show the temperature dependence of $\Delta$ (red cicles) and expectation from BCS theory (black line) with T$_c$=1.75 K.}}\label{fig3}}
\end{figure}

Our results establish that, in 2H-TaS$_2$, chiral CDW properties coexist with superconductivity. Possible consequences could be related to those found in the context of other superconductors with broken space inversion symmetry. For example, in a superconductor with broken space parity and strong spin orbit coupling (spin orbit coupling is strong in 2H-TaS$_2$ as discussed in Ref.\cite{Rossnagel07}), the two fold electronic spin degeneracy may be lifted, and spin structured Cooper pairs could appear \cite{Edelstein95,Schnyder10,Bauer10,Gorkov01,Samokhin08}. The form and size of the superconducting gap can be under such conditions peculiar, with nodes along some directions, but it can be also fully opened \cite{Suderow09,Mackenzie03}. In 2H-TaS$_2$, the observation of an opened somewhat anisotropic gap, similar to the ones found in related materials, shows that the influence of broken space parity in the superconducting properties is, if present, weak. Characterization of superconductivity in the other system of the series, 2H-TaSe$_2$\cite{Rossnagel07}, the observation of the vortex lattice in both compounds, and a closer look to superconductivity in related CDW materials, should bring further advance.

In summary, we have observed a chiral CDW in a metallic and superconducting compound, 2H-TaS$_2$, and provide measurements of its superconducting density of states. We could not establish an influence of broken space parity through a helical electronic structure on the superconducting density of states, but the mere presence of both phenomena in the same material, calls for more work in this and similar compounds.

We acknowledge discussions with P. Kulkarni, K. Machida, A.I. Buzdin, J. Flouquet and F. Guinea. The Laboratorio de Bajas Temperaturas is associated to the ICMM of the CSIC. This work was supported by the Spanish MICINN (Consolider Ingenio Molecular Nanoscience CSD2007-00010 program and FIS2008-00454 and by ACI-2009-0905), by the Comunidad de Madrid through program Nanobiomagnet, and by NES program of the ESF.

\section*{References}
%\bibliography{Lastbib5}

\end{document}